\documentclass[conference]{IEEEtran}

\usepackage{fancyhdr}
\usepackage[normalem]{ulem}
\usepackage[hyphens]{url}
\usepackage{microtype}
\usepackage{fixltx2e}
\usepackage{graphicx}
\usepackage[caption=false]{subfig}
\usepackage{authblk}
\usepackage[euler]{textgreek}
\usepackage{flushend}
\usepackage{subfiles}
\usepackage{amsmath}

\usepackage{siunitx} 
\usepackage{algorithm}
\usepackage[noend]{algpseudocode}
\usepackage{lipsum}
\usepackage{wrapfig}

\usepackage[bookmarks=true,breaklinks=true,letterpaper=true,colorlinks,linkcolor=black,citecolor=blue,urlcolor=black]{hyperref}
\def\BibTeX{{\rm B\kern-.05em{\sc i\kern-.025em b}\kern-.08em
    T\kern-.1667em\lower.7ex\hbox{E}\kern-.125emX}}

\fancypagestyle{firstpage}{
  \fancyhf{}

  \fancyhead[C]{\normalsize{
      \textbf{} \\ }} 
  \fancyfoot[C]{\thepage}
}  

\title{Hardware/Software Obfuscation against Timing Side-channel Attack on a GPU} 
\author{Elmira Karimi, Yunsi Fei, David Kaeli\\ECE Department, Northeastern University
Boston, MA 02115 USA\\\{elmirakarimi, yfei, kaeli\}@ece.neu.edu}

\begin{document}
\maketitle
\thispagestyle{firstpage}
\pagestyle{plain}

\begin{abstract}
GPUs are increasingly being used in security applications, especially 
for accelerating encryption/decryption.  While GPUs are an attractive platform in terms of 
performance, the security of these devices raises a number of concerns. One vulnerability 
is the data-dependent timing information, which can be exploited by adversary to recover the encryption key. 

Memory system features are frequently exploited since they create detectable 
timing variations. In this paper, our attack model is a coalescing attack, which 
leverages a critical GPU microarchitectural feature  - the coalescing unit. 
As multiple concurrent GPU memory requests can refer to the same cache block, 
the coalescing unit collapses them into a single memory transaction. The access 
time of an encryption kernel is dependent on the number of transactions. Correlation 
between a guessed key value and the associated timing samples
can be exploited to recover the secret key. 

In this paper, a series of hardware/software countermeasures are proposed to obfuscate 
the memory timing side channel, making the GPU more resilient without impacting performance.  
Our hardware-based approach attempts to randomize the width of the coalescing unit to 
lower the signal-to-noise ratio.  We present a hierarchical Miss Status Holding Register (MSHR) 
design that can merge transactions across different warps. This feature boosts
performance, while, at the 
same time, secures the execution. We also present a software-based approach to permute the 
organization of critical data structures, significantly changing the coalescing behavior 
and introducing a high degree of randomness. Equipped with our new protections, the effort 
to launch a successful attack is increased up to  $1433X \times 178X$, while also 
improving encryption/decryption performance up to $7\%$.
 
\end{abstract}

\section{Introduction}

Graphic Processing Units (GPUs) have become the accelerator of choice in a
number of compute-intensive applications.  GPUs achieve high throughput by
leveraging thread-level parallelism of applications on thousands of simple
cores.  The range of applications that can leverage GPU
acceleration spans deep learning models, computational physics, and security functions such as data encryption and signing~\cite{munshi2009opencl,Kaeli2015,nickolls2008scalable,nvidiamy1,nvidiamy2}.

CPUs are now offloading encryption workloads to GPUs, 
given their efficiency. Encryption/decryption operations are commonly 
present in security libraries and
they can benefit greatly from the high
performance provided by a
GPU~\cite{manavski2007cuda,iwai2010aes,biagio2009design,le2010parallel,
cohen2010gpu,szerwinski2008exploiting}. 
Iwai
et al.~\cite{iwai2010aes}, Li et al.~\cite{li2012} and
Schonberger~\cite{schon2011} each developed GPU implementations of the Advanced
Encryption Standard (AES) algorithm, demonstrating double-digit speedups on a
GPU compared to a CPU. However research has shown that GPUs are vulnerable to
attacks that leverage the physical properties associated with applications execution on the device,
a class of
vulnerabilities called side-channel
attacks~\cite{patterson2013vulnerability,karimi2018timing,jiang2016complete,
jiang2017novel}. Although some companies have developed secure encryption 
cores, these designs are specifically
implement  a single algorithm (e.g., AES)~\cite{akdemir2010breakthrough}.
Side-channel attack methods can be used to attack different table-based 
algorithms. Therefore we need a more comprehensive defense strategy.
Timing side-channel attacks have typically targeted the memory
hierarchy of a GPU.  If the attacker can measure the execution time of
an encryption, and with knowledge of the encryption algorithm's layout 
in memory,
the encryption keys can be recovered through correlation
analysis~\cite{karimi2018timing,jiang2016complete,jiang2017novel}.
RCoal~\cite{kadam2018rcoal}, has been 
proposed as a countermeasure that can change the warp size 
and thread allocation randomly. Unfortunately, this solution hurts
performance significantly due to reduced 
memory hierarchy efficiency, while only slightly improving security.

In this paper, we introduce a series of hardware and software countermeasures
against GPU coalescing attacks.  Given the fact that attackers
can use the relationship between execution time and number of memory accesses to
launch an attack, our approach attempts to decorrelate the measured timing
signal and memory access pattern.  In
most GPU memory systems, the {\em coalescing
unit} can dynamically merge multiple memory requests across different
threads into a single memory access, called a {\em transaction}.  Given the
organization of the data structures used during encryption, and their associated
key-dependent memory access behavior, the memory access pattern produces a
timing channel~\cite{jiang2017novel}.

To confuse the attacker, we first enhance the hardware by randomizing
the width of the coalescing unit to reduce the Signal-to-Noise Ratio (SNR) of
the timing sample. By modifying the width of this unit during program execution, we
can introduce randomness in terms of the number of memory transactions
generated.  In order to further reduce the SNR, we utilize
multiple coalescing unit widths across 
different cache lines.
By dynamically changing the width of coalescing unit for different
cache lines, we can introduce another 
dimension to the noise injected.  Since each {\em
Streaming Multiprocessor} (SM) in a NVIDIA GPU has its own coalescing unit, we
use a combination of different transaction sizes for each SM to introduce
additional noise to the timing channel. 
We also redesign the Miss
Status Holding Registers (MSHRs) to provide miss tracking across SMs, which not
only adds some additional noise to the timing channel, but also improves
performance.  In order to explore these changes in hardware, we utilize an
execution-driven GPU simulator.

For our software approach, we also focus on the memory model of the
GPU.  During AES
encryption/decryption we need to access several look-up tables 
stored in GPU memory.
We introduce a software-based 
lookup table hashing algorithm to obfuscate the associated access pattern
to one of these tables. This helps to hide the number of transactions observed
by the attacker through the measured timing sample.

The rest of this paper is organized as follows. Section 2 covers background on
GPU microarchitecture and describes a GPU-based coalescing attack. Section 3 provides an
introduction to our methodology.  In Section 4, we characterize the timing
channel present on a GPU, as well as the design of our hardware/software
countermeasures based on changing the coalescing unit width, modifying the MSHR
design and applying look-up table rotation.  Then we evaluate how well we can
thwart a timing attack in Section 5.  Finally, in Section 6, we conclude the
paper.

\section{Background and Related Work}

\subsection{GPU Architecture} 
In contrast to CPUs, GPUs typically provide many
cores. NVIDIA refers to these cores as Streaming Processors (SMs), while
AMD calls them Compute Units (CUs). 
In this paper we will use NVIDIA terminology
and provide a brief review of the Fermi GPU architecture. 
While Fermi is an older GPU architecture, 
it is used in our simulation infrastructure (GPGPU-Sim~\cite{GPGPUsim}), 
and most of the features
discussed in this paper are present in later generations of 
NVIDIA GPUs, as well as AMD GPUs.
So the attack strategies and defences proposed are relevant for a broad class of
architectures.  In this paper we use GPGPU-Sim based on the Fermi GTX480 GPU.
architecture, which has 15 SMs. Each SM includes 32 CUDA cores, 
a load/store(LD/ST)
unit, 4 Special Function Units (SFUs), a 64KB block of high speed on-chip memory
(L1+Shared Memory) and an interface to the L2 dcache. The L2 cache is
a unified shared cache across all the SMs.  Each SM can run two warps, 
with each warp running 32 threads in lockstep. Warps are scheduled by a warp
scheduler. Detailed configuration information is provided in 
Table~\ref{table:1}. 

\begin{table}[h!] 
\caption{The Fermi GTX480 architecture.}
\centering 
\begin{tabular}{ | m{10em} | m{2cm}| m{1cm} | } 
\hline Number of SMs & 15  \\ \hline Number of Threads per Warp & 32  \\ 
\hline
L1 Data Cache Size & 48KB \\  \hline L2 Cache Size  & 768KB \\ \hline
Transaction Size & 64B \\  \hline Memory Bus Width & 258B \\ \hline 
MSHR
Entries & 32 \\ \hline 
\end{tabular} 

\captionsetup[table]{skip=30pt}
\label{table:1} 
\end{table}

\textbf{Memory Coalescing:} Threads
within a warp run in a Single Instruction Multiple Thread (SIMT) fashion.  As a
result, each memory instruction in a warp will produce 32 memory requests,
which are sent to the LD/ST unit.  Each LD/ST unit includes a coalescing unit 
that inspects the addresses generated across the warp and collapses
them into a minimum set of memory transactions. 
The size of each transaction is equal to one cache line. Accesses are 
non-blocking, such that memory
can handle multiple requests concurrently.
Wang et. al~\cite{wang2018intra}
proposed an intra-cluster coalescing unit to merge memory requests from different
SMs in a cluster. 

\textbf{Miss Status Holding Register Functionality:} 
Every SM
has a private L1 cache.
On an L1 read miss, an entry in the Miss Status
Holding Registers (MSHRs) will be allocated for tracking the miss.  
Each outstanding request to the L2 cache is tracked by an MSHR entry.
Once the L2 cache responds, the MSHR entry will be released. 
Each MSHR entry can 
be shared by multiple accesses in an SM in order to avoid redundant entries,
though redundant entries from across different warps may cause a performance loss.
Tuck et.
al~\cite{tuck2006scalable} described sharing MSHR entries
across L1 banks, improving bandwidth management to L2. 

\subsection{Side-channel Attacks}
Side-channel attacks exploit information recorded during execution
of encryption/decryption. There are many types of attacks and countermeasures in different levels of the design~\cite{keshavarz2017privacy,keshavarz2018sat,keshavarz2017design,vosoughi2019combined,vosoughi2019bus,vosoughi2019leveraging}.
The memory hierarchy is one of the most common targets for the
attacker. There
are two main types of memory attacks: access-driven and time-driven
attacks. In an
access-driven attack, the attacker exploits the principle
of cache sharing to
observe the victim's memory access behaviour. The adversary 
intentionally creates contention on the shared cache resource 
to be able to infer if certain addresses have been accessed by the 
victim.  Alternatively, a time-driven
attack is based on measuring the execution time of an encryption/decryption and
establishing a correlation between this timing value 
and an embedded secret.  

Most previous studies of side-channel attacks, as well as 
attack countermeasures, have targeted CPU devices. Bernstein~\cite{bernstein2005cache} 
proposed a simple class of timing attack
strategies that leverage the layout of data structures in the memory hierarchy
of a CPU and used the correlation between the data structure index and the
measured time to recover the encryption key.  Tromer et
al.~\cite{tromer2010efficient} describe several access-driven attacks 
and countermeasures against such timing. These methods attempt to obscure the timing channel by
introducing time constant algorithms, masking the access pattern by disabling
shared memory or performing
random permutations and preloading. Both software and
hardware approaches have been used, but the main problem with most of 
these obfuscation techniques
is that they come with significant performance costs, rendering them infeasible
or costly to implement.  

Given the growing
popularity of GPU platforms, it is important to thoroughly evaluate vulnerabilities of
these high-performance devices, as well as employ effective countermeasures
against various attacks.  The most important difference between CPUs and
GPUs is the source of the execution time variation. In CPUs,
the variations in timing are the result of cache misses (or hits), providing a clean timing channel. 
While on a GPU, instead of tracking cache hits and misses, timing variations
are a result of memory request management by the GPU's 
coalescing unit. While   
the degree of thread-level concurrency is very high on a GPU, the noise
associated with this concurrency is not enough to obscure the timing
channel~\cite{karimi2018timing, jiang2016complete, jiang2017novel}.  The
attacker generally only observes a single timing value associated with 
the execution
of all the threads in a kernel.  Although, the correlation between 
the execution
time and the memory access pattern becomes 
weaker as the degree of concurrency is
increased, the GPU memory reference pattern is mainly a function of the plaintext
input and encryption/decryption 
algorithm. An attacker can track data access patterns using timing. Even if
the programmer chooses to store the encryption data structures in shared memory,
a partition of memory only accessible to each Streaming Processor on the device,
a successful attack is still possible.  The good news is that the same memory features
that provide attackers with an attractive attack surface, can also be used as a
form of protection.

\subsection{Baseline Attack} Memory coalescing is a common feature that has been
used by GPU designers to improve memory throughput.  This mechanism also
provides a strong timing channel, since the effectiveness of the coalescing unit
is a function of the memory address pattern generated across threads.  An
attacker can use this timing information to quantify the number of transactions,
and eventually derive the encryption key by computing the correlation between
the measured time and number of transactions. This attack scenario was first
identified by Jiang et al.~\cite{jiang2016complete}, where the attacker remotely
sends a large number of random {\em plaintext} inputs to the victim. The victim
uses a GPU to encrypt the data and the attacker finally collects the resulting
{\em ciphertext} and the execution time.  The attack is repeated many times
(typically 100K to 1M samples are collected). 

\begin{equation} 
\label{eqs} 
C_{j}= T_4[t_{i}] \oplus k_{j}:
0<j<15,0<i<15
\end{equation} 

This attack method is adopted as our baseline methodology.  Based on this
scheme, we will introduce randomness during encryption to obfuscate the timing
channel.  Jiang et al. described a T-table implementation of AES encryption (ECB
mode).  We implement the same scheme in CUDA, following the OpenSSL 0.9.7
library implementation. The CUDA kernel takes 16 bytes of plaintext and 16 key
bytes, and outputs ciphertext. The key bytes are expanded to 160 bytes, because
in total AES has 10 rounds. First, the 16-byte plaintext is XORed with the first
round key.  Then during the next 9 rounds, a combination of the four T-table
lookups ($T_0$, $T_1$, $T_2$ and $T_3$) are XORed with the key. For the last
round, another T-table, $T_4$, are performed. As there is no 
\textit{MixColumn} operation, the ciphertext
can be derived using Equation~\ref{eqs}.

\begin{figure}[h] 
\centering 
\includegraphics[width=0.4\textwidth]{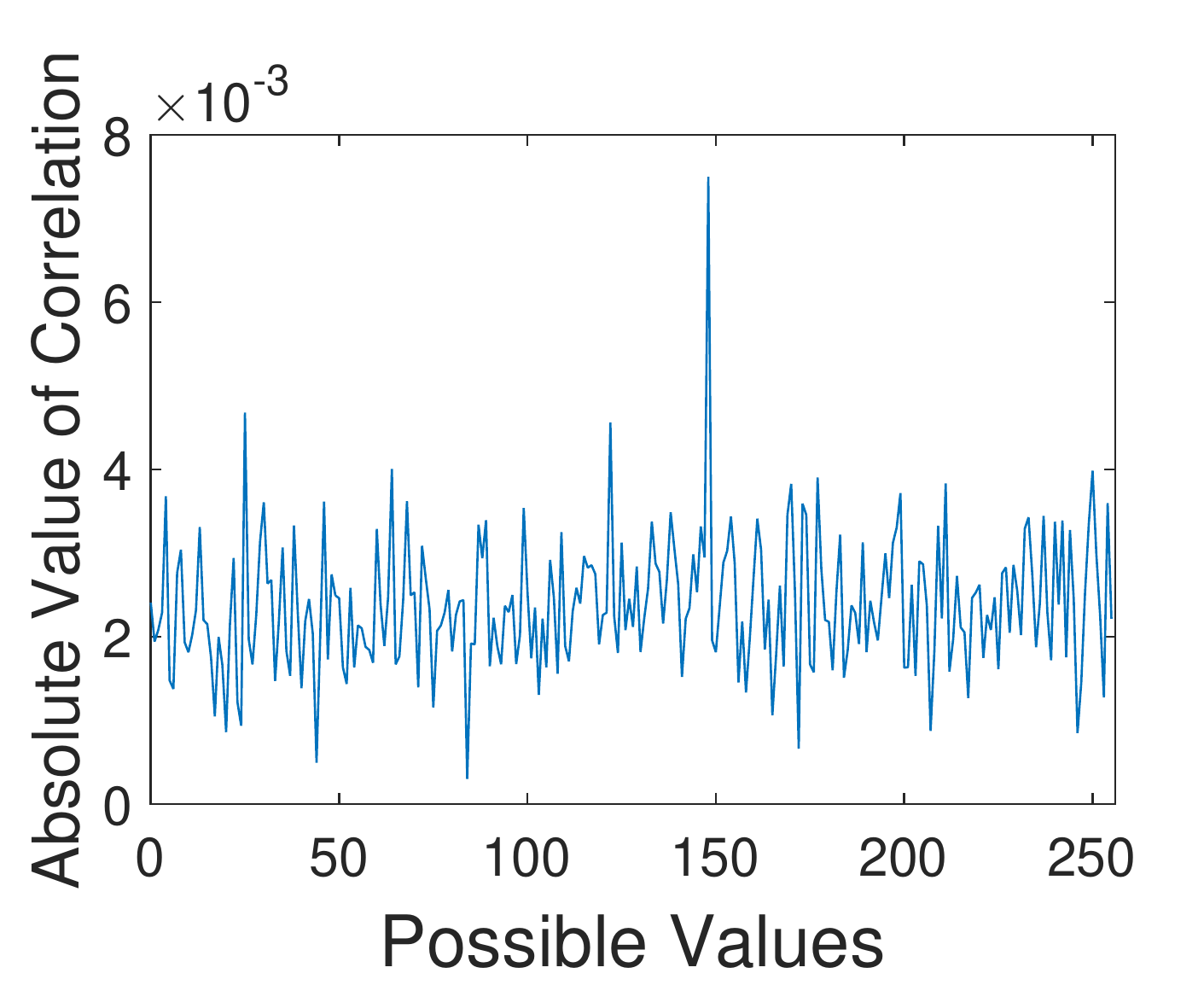} 
\caption{The
absolute value of the correlation for possible values of $k_5$.} 
\label{attack}
\end{figure}
In our CUDA implementation, the entire encryption is implemented in a kernel.
Each thread in the kernel takes 16 bytes of plaintext, performs an independent
encryption, and produces 16 bytes of ciphertext. A warp contains 32 threads, so
32 encryptions are performed together. Each T-Table has 256 4-byte elements
(1KB in total). The T-table is stored in global memory.

In the attack scenario, each key byte is guessed (from the values 0 to 255). By
knowing the ciphertext of each sample and using Equation~\ref{eqre}, the
attacker can produce all indices ($t_i$).
Equipped with a coalescing unit, the 32 concurrent memory requests 
result in a smaller (and computable) number of memory transactions.
The attacker only needs to know the width of the coalescing unit. 
For example, if the transaction size is 32 bytes, each memory transaction would
contain 8 consecutive addresses.  To recover the transaction index, the attacker
simply shifts the original index by $log_2 8$ (i.e., 3).  After determining the
number of transactions based on the cipherthext, the attacker can calculate the
correlation between the measured execution time and the predicted number of transactions under a key byte guess.

For a key byte, there are 256 possible values.  The attacker can calculate the set of 
256 correlation values using the execution time and the number of transactions
for each key guess. The value with the highest correlation between the execution
time and the number of transactions will be associated with the correct key
guess. The attacker will repeat this process for the
other 15 bytes of the key and find the highest correlation for each.
Figure~\ref{attack} shows the correlation
of all possible values of $k_5$, using 500,000 timing samples.

\begin{equation} 
\label{eqre} 
t_{i}=T_{4}^{-1}[C_{j} \oplus k_{j}] : 0<j<15,0<i<15 
\end{equation}

\section{Our Methodology}

Typically, the degree of difficulty associated with an attack can be measured by
the number of timing samples needed to recover the key.  An ideal
countermeasure increases the number of samples required for the attack, without
hurting execution performance.  Kadam et al.~\cite{kadam2018rcoal} changed the
number of threads that are grouped together in a warp randomly. While
introducing randomness to help protect a device, GPUs are designed to execute
regular (e.g., dense matrix) operations, where the execution time
is well behaved. So
reducing the warp size in order to introduce noise in the timing channel should
greatly impact performance.

We introduce a series of hardware/software countermeasure based on commonly design
principles used in most GPU memory systems. Given the fact that the attacker can
use the relationship between time and the 
associated memory access pattern, our approach
attempts to reduce the correlation between number of memory accesses and the
measured timing signal. 

\subsection{Obfuscation in Hardware}
\subsubsection{Randomizing The Width of the Coalescing Unit} 
The transaction size of coalescing unit is
equal to the size of an L1 cache line. The address range of the coalescing
unit is always fixed. Across the threads in a
warp, each  address pattern will result in a predictable number of cache line
accesses. In order to confuse any attacker who is tracking memory behavior, we
propose an obfuscation approach that dynamically
modifies the coalescing unit. By changing
the range of addresses that result in a transaction, we render the execution
time associated with a memory address pattern as non-deterministic.  For
example, if the cache line size is $l$ bytes, the basic coalescing unit maps
addresses from the range $0$ to $l-1$ into the first cache line, but with our
coalescing unit, the width of coalescing unit is randomly selected 
to be a  
fraction of the total cache line size.  This will introduce a source of noise,
and by randomly changing the number of transactions, this
will significantly increase the number of executions required to launch a 
successful
coalescing attack.

We start by 
randomizing the width of the coalescing unit based on a normal probability
density function. In this step, the width of the 
coalescing unit is randomly chosen,
but the value
is fixed for each individual cache line. To inject more noise, 
we can use different coalescing widths for different cache lines. 
One simple way to do this is to generate a random width for 
every cache line. Given that an L1 cache has a large number of cache lines, 
generating a random value for the width to be used for every line 
would be costly. In order to reduce the overhead,
we generate a limited number of random widths, for example 16, 
and use this set of widths for every 16 cache lines, and then 
repeat the pattern through the whole cache.
This approach introduces additional noise
into the timing channel, and makes it significantly harder for an attacker to
recover the key.

\subsubsection{A Hierarchical MSHRs for Increased Obfuscation Strength} 
MSHRs help to manage memory bandwidth when encountering L1 misses. GPUs provide
a private set of MSHRs for each SM. Kernels tend to fully utilize 
all SMs in order to exploit the highest degree of parallelism available on a
GPU. As a result, there can be a varying number number of misses
experiences across different SMs to the unified L2, which results
in a performance loss due to L2's limited bandwidth and the number of 
MSHRs per SM available. To provide further obfuscation, while also 
considering performance,  
we present a hierarchical MSHRs design, where at the
first level, each SM has its own
set of MSHRs.  Then at the second level, the MSHR 
entries are shared across all SMs.
This can benefit both the
security and performance of the system. In terms of security, a
hierarchical
MSHRs
will reduce the number of L1 misses from different SMs and hide a portion
of the latency, which introduces additional noise into the timing 
channel. Our
proposed design should also improve memory performance as compared to 
current GPU-based MSHRs. 


\subsection{ Obfuscation in Software}

Changing the T-table mapping in the cache is a commonly
proposed countermeasure which attempts to 
randomly permute the address mapping using either
software or hardware. Wang et al.~\cite{wang2007new} proposed novel
cache design named {\em RPcache}, which can only defeat access-driven attacks.
Brickell et al.~\cite{brickell2006software} considered obfuscation
that employed a software solution, though a fixed
\begin{figure}[h] \centering
\includegraphics[width=0.4\textwidth]{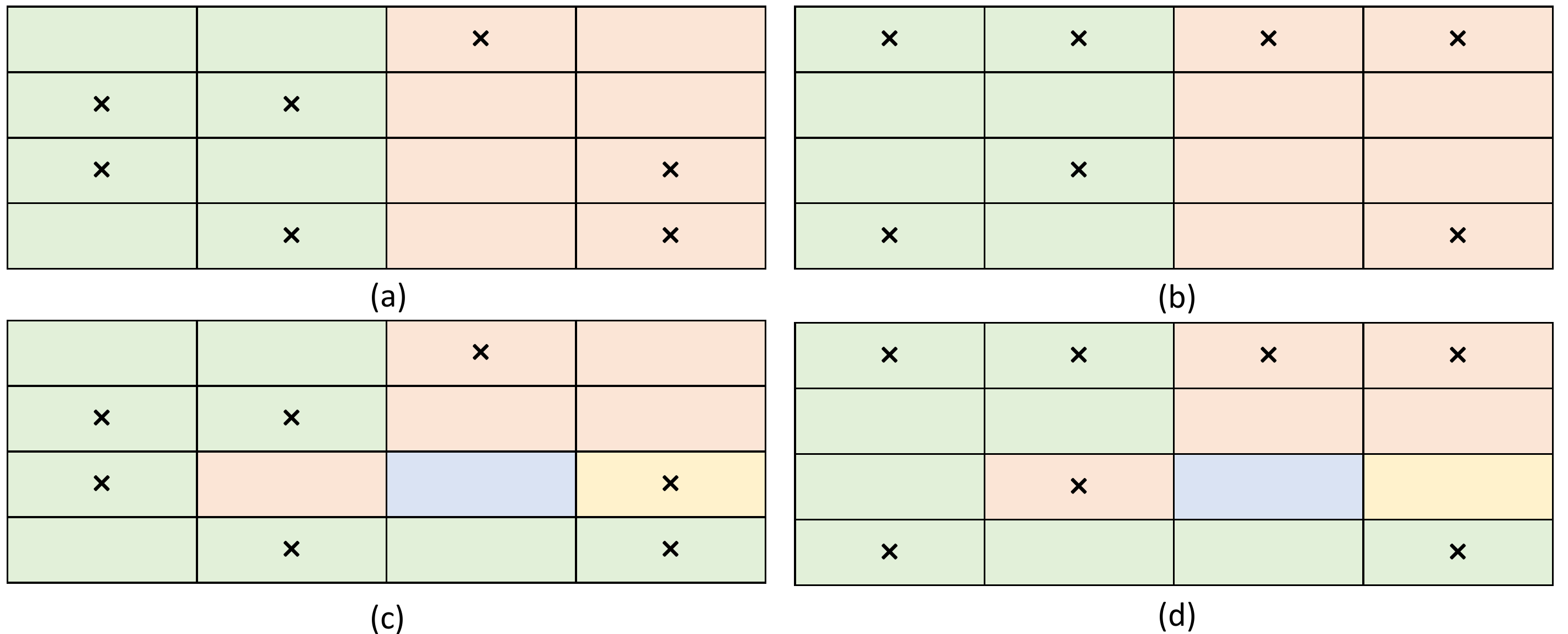} 
\caption{Hardware and
Software Obfuscation using: a) a fixed coalescing unit width,
b) a fixed coalescing unit
width with rotated accesses, c) a dynamic coalescing unit width 
(for row 1 to 4 the
widths are 2, 2, 1 and 4, respectively), and (d) a dynamic coalescing 
unit width combined with rotated
accesses.} 
\label{dynwidth} 
\end{figure}
permutation only obfuscates the timing channel temporarily 
until the attacker figures out the permutation algorithm.
In this paper we propose a GPU-oriented random rotating algorithm
that works by rotating T-table addresses, changing the permutation
over time. The frequency of applying rotation can be controlled
to minimize the overhead, while providing the right level
of obfuscation.  Dynamic
rotation also can exploit
the parallelism of a GPU to prevent performance loss.  

Our approach effectively disassociates elements stored in a cache line and 
re-associates them with new elements in other
cache lines. As an example, 
suppose that we have $n$ elements in the T-table that will be stored in
$l$ continuous cache lines and each cache line contains $m$ elements. 
Figure ~\ref{dynwidth}(a) shows a simple case in 
which the look-up table has 16
elements and each cache line can store 4 elements. As we can see, this is a 
two-dimensional structure, where the rows represents cache lines. In this
example, we have 7 random accesses, where 
the number of accessed cache
lines (i.e., rows) is 4. The attacker can
use this number to launch an attack by finding the
correlation between the execution time and the number of cache accesses. One approach to
obfuscate the
number of rows accessed is to change the mapping of each entry, 
column-by-column, as shown in Figure~\ref{dynwidth}(b),
where columns 1, 2, 3 and 4 are shifted by 2, 3, 0 and
1,
respectively
This rotation changes the
access pattern as shown,
resulting in changes to the row accessed from 4 to
3. 

Figure~\ref{dynwidth} also shows the impact of obfuscation by
changing the  
coalescing unit width, combined with rotation in the T-table. 
In Figure~\ref{dynwidth}(a), we show results using a 
fixed coalescing unit width ($=2$) with different colors. In Figure~\ref{dynwidth}(b), we
combine a fixed coalescing unit width with the
T-table rotation (as described before). 
Figure~\ref{dynwidth}(c) 
shows results for dynamically changing the width
of the  coalescing unit
for different rows. Figure~\ref{dynwidth}(d) shows results when dynamically
changing the
coalescing unit width, combined with
T-table rotation. The number of rows accessed is 6,
5, 5 and 4 for cases a-d, respectively.

\section{Hardware/Software Countermeasures}
In this section, first we present a series of hardware countermeasures
based on a redesign of the GPU's coalescing unit. Then
we provide details of our
hierarchical MSHR design.  Our solution 
improves performance and provides further
protection. Then we discuss our software approach, which randomly
rotates columns of the T-table in the GPU's global memory.

\subsection{Randomizing the Width of the Coalescing Unit} 
As discussed earlier,
we use GPGPUsim to evaluate a timing attack of an AES running on a Fermi GPU. In this
simulator, we set the memory transaction size to 64 bytes, which is equal to L1
cache line size.  To study the effects of changing the width of the coalescing
unit on the execution time of a kernel, we first develop a simple kernel to
learn the behavior of the coalescing unit and then design the  countermeasure
based on our observation.  

\subsubsection{benchmarking Observations} 
\begin{figure}[h] 
\centering 
\includegraphics[width=0.5\textwidth]{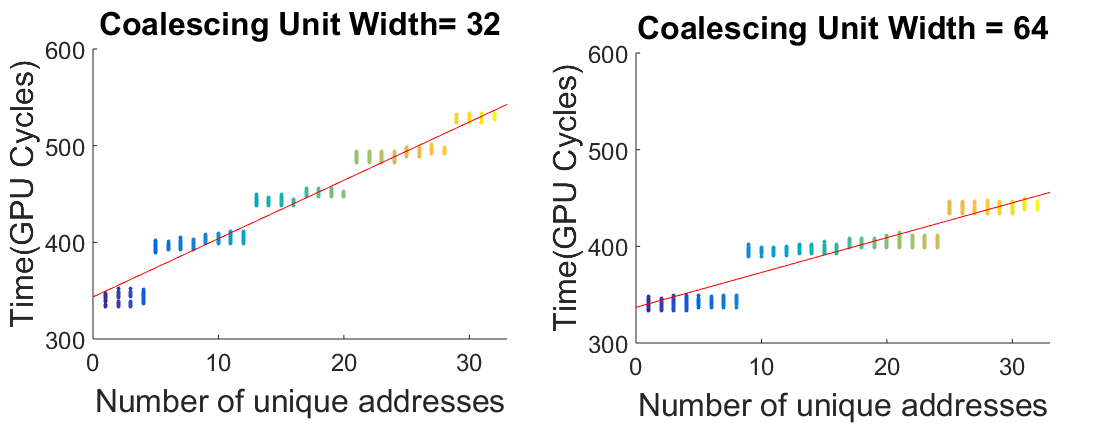}
\caption{Execution time of accessing 1 to 32 unique address with different width
of coalescing unit} \label{varall} 
\end{figure}

We developed a microbenchmark
kernel to study the impact of changing the width of the coalescing unit.
The kernel loads
an array of floats from global memory, where 
each warp generates 32 loads.  We set
the array indices to access \textit{n} unique addresses.  We modify this
pattern, varying the value of \textit{n} from 1 to 32 over each kernel, for 32
threads. We vary the width of the
coalescing unit, setting this value to 8, 16, 32
and 64 bytes.  Our design space includes $32\times4$ configurations.  Each
microbenchmark kernel is run 10,000 times per configuration. We collect the
corresponding timing samples for each configuration, $128\times10^4$ timing
samples in total.  Figure~\ref{varall} presents the measured execution time,
versus the number of unique addresses, for coalescing unit widths of
32 and 64 bytes. We added a random noise to the simulator's timing to simulate the real hardware.
As the results show, the execution time is linearly proportional to the number
of unique addresses, and changing the width of coalescing unit changes the slope
of this line.  Using \textit{regression analysis}, we can construct
a linear equation
that expresses the relationship between the time samples and the number of
unique addresses. The coefficients of this equation are found using a
least-squares approximation method, which minimizes the distance between the
actual timings and the expected ones.  In equation~\ref{equ1}, $t_e$, $n$ and
$\epsilon$ are the execution time, the number of unique accesses and the error
due to noise, respectively.  We use regression analysis to find $\beta_1$, which
represents the slope of the line and $\beta_0$ that is a constant.

\begin{equation} \label{equ1} t_e=\beta_1n+\beta_0+\epsilon \end{equation}  

The attacker can use the fact that the time measured
is linearly proportional to the
number of unique accesses, and can be inferred from the execution time. But
when varying the width of the coalescing unit, we observe a less tractable
relationship between time and the memory addresses.  This confirms that the
design of the coalescing unit can impact the leakiness of the GPU.  
One metric that can be used to measure the 
{\em leakiness} of an architecture is SNR. A higher
SNR is desirable for the attacker. Therefore, to protect a GPU from a
side-channel attack, we should try to reduce the SNR. The SNR for the linear
model produced by our regression analysis (Equation~\ref{equ1}) is calculated by
Equation~\ref{equ2}, where $\sigma^2_s$ and $\sigma^2_\epsilon$ represent the
variance of the signal and the noise.  

\begin{equation} 
\label{equ2} 
SNR= \dfrac{\sigma^2_s}{\sigma^2_\epsilon}=
\dfrac{\beta^2_1\sigma^2_n}{\sigma^2_\epsilon} 
\end{equation} 

If the access pattern and noise remain constant, increasing the slope of the
line results in a higher SNR.  In other words, $\beta_1$ captures how likely the
attacker can distinguish addresses $i$ and $i+1$ when she records the
corresponding execution times ($t_e(i)$ and $t_e(i+1)$). The first four rows of
Table~\ref{table:2} show the characteristics of the linear model while varying
the width of the coalescing units. As we can see, by increasing the width of
the coalescing units, 
$\beta_1$ is reduced, which results in a lower SNR based on
Equation~\ref{equ2}.  To calculate the SNR using Equation~\ref{equ2},
$\sigma^2_\epsilon$ is calculated in Equation~\ref{equ3}. 

\begin{equation} 
\label{equ3} 
\sigma^2_\epsilon= \frac{\sum_{i=1}^{m}
(t_e-(\beta_1n+\beta_0))^2}{m}
 \end{equation}
 
  \begin{table}[h] \centering
  \caption{Regression analysis characterization.}
\begin{tabular}{ l c c c } \hline  width of coalescing unit & $\beta_1$ &
$\beta_0$  & SNR  \\ \hline 8 & 19.463 & 346.2 & 1.8656 \\ \hline 16 & 11.42 &
344.3 & 0.9429 \\ \hline 32 & 6.032 & 343.6 & 0.4494\\  \hline 64 & 3.601 &
336.9 & 0.1940\\  \hline Fixed Random Width & 4.426 & 337.3 & 0.0985 \\ \hline
Dynamic Random Width & 4.087 & 336.6 & 0.0050 \\ \hline 
\end{tabular}
 
\label{table:2} 
\end{table}

The results of our micro-benchmarking exercise illustrate how changes in the
width of coalescing unit affect the SNR.  Using the maximum width of the
coalescing unit, the SNR is still high due to the lack of noise. One parameter
to represent the noise is the standard deviation (STD) of the measured execution
time. A lower STD means that the timing samples are less noisy. In our
observations, when running our microbenchmark kernel 10,000 times, the STD is
very low, meaning that the execution time is still highly correlated with the
number of unique accesses. So the attacker can filter out noise by running a
reasonable number of samples. To introduce more noise into the system, we
randomly change the width of the coalescing unit to increase
$\sigma^2_\epsilon$, and eventually, reduce the SNR. This randomness can be
fixed and dynamic, which will be described next.

\subsubsection{Using a Fixed
Random Width for the Coalescing Unit} 
\begin{wrapfigure}{r}{0.2\textwidth} 
\begin{center}
\includegraphics[width=0.2\textwidth]{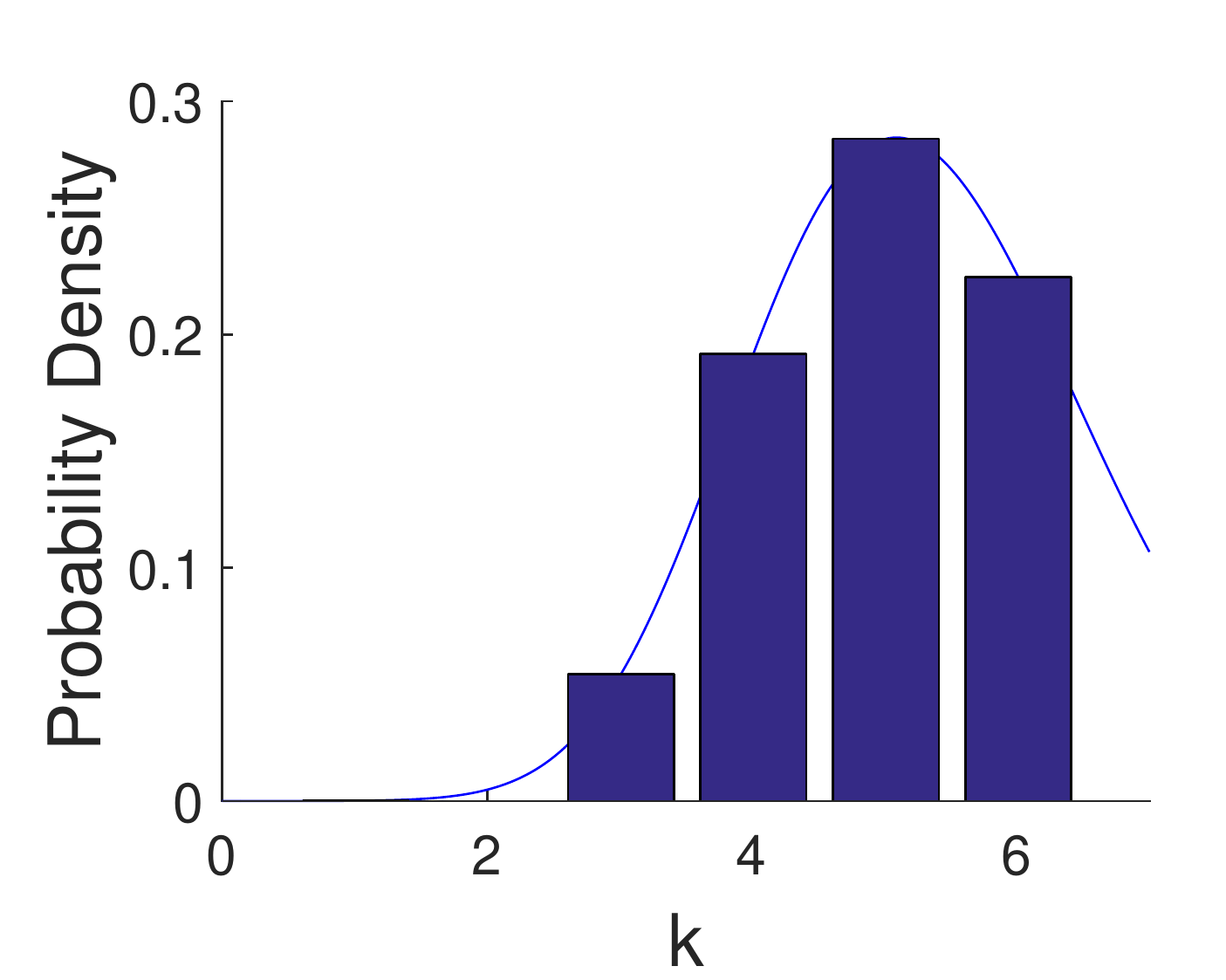} 
\end{center} 
\caption{Probability density function, while changing  
the width of the coalescing unit.} 
\label{spacesk} 
\end{wrapfigure}

Introducing fixed randomness means that during
each kernel run, the width of the
coalescing unit is randomly chosen, but the width is fixed
for all cache lines. Here, we have used 4 different coalescing unit 
width values
as our reference set. The width of coalescing unit is $2^k$ bytes, where $k$ is
varied from 3 to 6. The maximum width of the coalescing unit (64 bytes) 
is equal to the width of an L1 cache line.
The total number of timing traces collected is still 10,000
and the probability density function (PDF) representing the coalescing unit
width is a skewed normal distribution, with a mean $k=5$, as shown in
Figure~\ref{spacesk}.  By using this distribution, the probability of choosing a
larger transaction size is much higher (e.g., for only 500 out of 10,000 runs,
we see 8-byte transactions).  If we run 10,000 samples while selecting
a  random width for the
coalescing unit, the SNR is reduced
(as shown in Table~\ref{table:2}) due to the increased noise.  

\subsubsection{Using a Dynamic Random Width for the Coalescing Unit:} 
As we observed,
randomizing the width of coalescing unit increases the amount of noise and
reduces the SNR. In regression analysis, the variance of the noise is equivalent
to the sum of the squared distances between the predicted and actual
values.  Randomly changing the width of the coalescing unit increases this
distance, resulting in more noise.  

\begin{figure}[h] 
\centering
\includegraphics[width=0.3\textwidth]{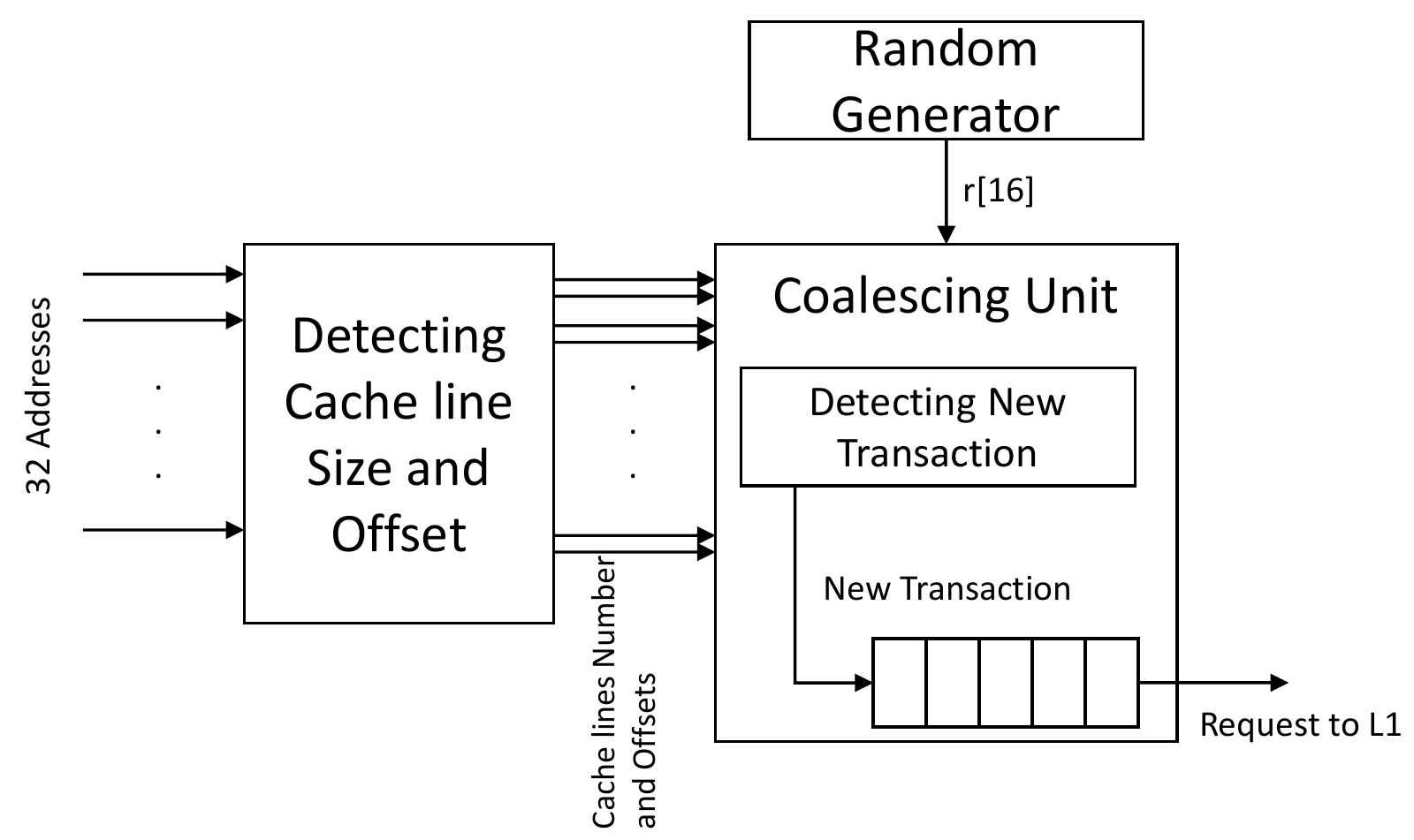} 
\caption{Redesign
Coalescing Unit for Dynamic Random Width} \label{updatedcoal} 
\end{figure} 

But
we still utilize a fixed random value selected from a reference set. Another
level of randomness would be to use different widths for different cache lines.
But generally speaking,
supporting a unique random width for every cache line has a huge
overhead. Therefore, we choose a limited number of cache lines and produce $n$
different width values and repeat the random width pattern for every $n$ cache
lines. Here, we choose 16 cache lines due to the
T-table size, which occupies 16 cache
lines. A high-level design for this functionality is shown in 
Figure~\ref{updatedcoal}. 

In this design, a pseudo-random number generator generates 16 
values during each kernel run, $r[16]$, 
from the set ${1,2,4,8}$, recognizing that each cache
line width should be split into $r[i]$ parts that we call
\textit{subtransactions}. Once a request is issued by a warp, 
the cache lines and
offsets within the cache lines are calculated for all 32 requests. At this
point, a traditional coalescing unit would check each request to see 
if there is a memory request being processed for this cache line being
issued by another thread. 
If a match is found, 
the current request will be coalesced and checking will move
to the next request. 

In our proposed 
design, when the coalescing unit receives a 
request and finds an access referencing the same cache block  
in the ordered memory request queue, 
it will first check the number of subtransactions within the cache line
based on the $r$ array entry, indexed as follows: $r[cacheline
number\%16]$. Then the width of each subtransaction 
will be calculated as:  $\frac{64}{r[cacheline
number\%16]}$. We refer to this width as the {\em subtransaction width}.
The coalescing unit will resolve the subtransaction number 
of the current request (0 to $r[cacheline
number\%16]-1$) based on the request's offset,
and if there is a queued request with the same cache line 
and subtransaction number for the same cache line, 
the current request
can be coalesced.  Otherwise, a new transaction will be sent to L1. 
While this new transaction is possibly unnecessary given that there is 
already an outstanding transaction for the same cache line underway, the
added time will help obfuscate this timing channel. We provide the SNR values of this case in
Table~\ref{table:2}. Using this method makes the design more resilient to the attack due to randomly chaning the pattern in every single kernel run.

\subsection{Hierarchical MSHR Design}
As we discussed earlier, each SM in our architecture has a set of 
32 MSHRs for
tracking L1 misses. One issue with this microarchitecture is that threads across different
SMs could request the same data from L2, generating redundant requests and 
producing a bottleneck at L2. This issue can significantly
impact performance when many SMs are making requests to L2. 
In order to avoid this issue, we introduce a second level
of MSHRs.  This set of
MSHRs is unified and shared across all SMs, with each MSHR having 
32 entries. In this
design, each L1 miss first allocates an entry in 
the MSHR local to the SM issuing the 
request to L2. These requests arrive at the second-level MSHRs, and if there
are no outstanding requests issued by other SMs for the same L2 cache line
in the unified MSHRs, the request
will be sent to the L2 cache. This design will 
increase performance whenever the competition for L2 access by multiple
SM is high. Although a unified MSHR set is a shared hardware structure across SMs, 
it cannot be an attack surface. First of all, in most GPU applications, the GPU
resources are fully
utilized such that 
the attacker would not be able to run their application concurrently with victim's. Second, a 
kernel would be run on several SMs and all of these SMs would send L2 requests
via the MSHRs. Therefore, even though the attacker can get an SM to observe the 
requests of a unified MSHR, it is very challenging to track the requests coming 
from different SMs that are executing concurrently. 

While our MSHR design 
has the potential of improving performance, the
second-level unified MSHRs will introduce another source of obfuscation.  
A hierarchical MSHR set can merge and collapse different incoming requests
within a range of 128 bytes(L2 cache line size).
This architecture will lower the SNR due to hiding the latency of L1 misses. In
Equation~\ref{equ1} we have modeled the time as a linear function of the 
number of
transactions. Suppose that we have $\beta_{10}$ L1 misses
resulting from $\beta_1$ memory accesses.
We can rewrite Equation~\ref{equ1} as Equation~\ref{equ5}, where $h$ is the
hit time and $m$ is the miss time for the L1 cache and $m=m_0h$ .  

\begin{equation} 
\label{equ5}
t_e=(\beta_{10}m_0+(\beta_1-\beta_{10}))hn+\beta_0+\epsilon 
\end{equation} 

When
we use a hierarchical MSHR set, there is some possibility that we
will reduce the miss penalty for
some L2 requests. This nondeterministic behavior will
add noise to $m_0h$ by 
reducing the miss time to $m_{noise}h$, which is in the range of
the time required to service the outstanding miss. 

If the requests can be merged with probability
$p_{m_0}$, then Equation~\ref{equ5} can 
be rewritten as Equation~\ref{equ6}.
In Equation~\ref{equ61}, $\epsilon_0$ is the noise generated
using our hierarchical
MSHR set. 

\begin{equation} \label{equ6}
t_e=(\beta_{10}m_0+(\beta_1-\beta_{10}))hn+\beta_0+\epsilon-p_{m_0}(m_0-m_{noise})\beta_{10}nh
\end{equation} 

\begin{equation} \label{equ61} t_e=
(\beta_{10}m_0+(\beta_1-\beta_{10}))hn+\beta_0+\epsilon+\epsilon_0
\end{equation} With this architecture the SNR will be lowered due to adding
$\epsilon_0$ as: 
\begin{equation} \label{equ7} SNR=
\dfrac{\beta^2_1\sigma^2_n}{\sigma^2_\epsilon+\sigma^2_{\epsilon_0}}
\end{equation} 

\subsection{GPU-Oriented Random Rotation} 
During a GPU
timing attack, the only information available to the attacker is the 
execution time and kernel inputs. The attacker attempts
to correlate the execution time with predicted number of transactions under each key guess in view of the coalecsing width. Another
way to decorrelate these two parameters is 
to change the access pattern of the
threads to memory, which results in a nondeterministic
number of transactions. In
order to change the access pattern, 
we should study the GPU's memory model.  Based on how 
we map data in to the GPU memory, we can introduce nondeterminism to the access pattern.
\begin{figure}[h] 
\centering 
\includegraphics[width=0.4\textwidth]{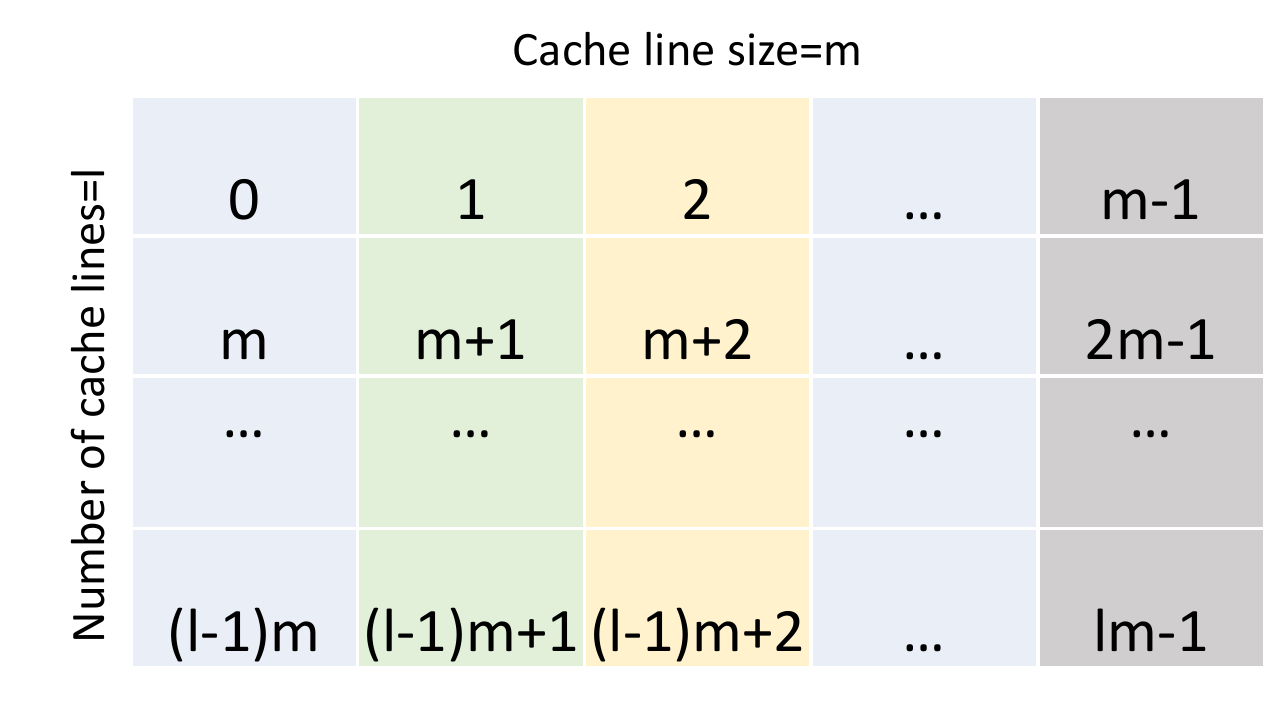}  \caption{General structure of a lookup table with size $n=ml$ in cache,
where the cache line can hold $m$ elements.} 
\label{struct}
\end{figure}

In an AES encryption/decryption baseline attack,
5 T-table are copied from the CPU memory to the
GPU memory. As the attack targets the last round 
of the encryption by tracking accesses to $T_4$, we will
focus on mapping of this table to GPU memory. While we focus
on accesses to $T_4$ due to their observability,
we can apply rotation to any look-up table. A
look-up table is a one-dimensional array 
of size $n$.  If we assume the  
whole array can fit in the L1 cache, and each cache line can store
$m$, then the table would occupy $l$ lines,
where $l=\left[\frac{n}{m}\right]+1$.
Figure~\ref{struct} shows the general structure of this look-up table in cache.
Whenever threads within a warp make a request to memory, the coalescing unit combines
the requests within a cache line (without randomizing the width of coalescing
unit) and will access the whole cache line. Therefore, the number of accessed
cache lines depends on the row-wise location of the request. To introduce
randomness to the number of transactions in this structure, we 
can dynamically change the mapping of an address to the cache.

\begin{algorithm} 
\caption{Random Rotation}\label{al1} 
\begin{algorithmic}[1]
\State $m \gets SizeofCacheline()$ \State $n \gets SizeofTtable$ \State $l \gets
n/m+1$ \For {$\textit{i} = 0:m-1$} \State $r \gets UniqueRand()$ \For
{$\textit{j} = 0:l-1$} \State $T_4[((j+r)\%l)m+i] \gets T_4[jm+i]$.  \EndFor
\EndFor 
\end{algorithmic} 
\end{algorithm}

An example of mapping the look-up table is when
we want to map element $(i,j)$ to $(i+i_0,j+j_0)$, which is in the
range of $[m,l]$. But as discussed, 
the coalescing unit accesses the whole row and produces a deterministic memory behavior.
But if we use a {\em column-wise random rotation}, we can introduce
noise into the timing channel. Algorithm~\ref{al1} provides
pseudocode for rotation algorithm. Using this scheme, each column will
circulate based on a unique randomly chosen
number between $[0,l-1]$, providing the highest level of diffusion across
the cache as possible.

The rotation can be performed on either 
the CPU or the GPU side. The main 
drawback of performing this transformation on the CPU
side is that the rotation would be performed only one time, 
since it is too expense to repeat the memory transfer from the CPU
to the GPU.  When performed on the GPU side, 
this not only eliminates data transfer overhead, but also
our rotation can be done in parallel 
on the many cores of the GPU.
As a result, the performance overhead is much smaller, thus we use a GPU
implementation, and can frequenty run a table rotation. The question
of how often rotation is necessary is discussed next.

The last round of AES is the target for our attack. This round 
involves a number of 
look-up table accesses to table $T_4$. The size of the table 
is 1K entries, occupying 16 64B cache lines. Suppose that we rotate
columns in 
the T-table every 100,000 kernel executions. So every 100,000 
iterations, the algorithm~\ref{al1} is executed. Each thread is 
responsible for rotating a column of 4B elements, 
16 threads in total. Then the AES encryption will be started. Then,
the probability of having the same number of transactions is 
significantly decreased, which reduces the correlation 
between execution time and number of transactions significantly.


\section{Countermeasure Analysis}
 
One approach to defend against a timing
attack would result in an increase in the  
number of timing samples required to launch a successful attack. 
As we discussed earlier, the attacker
uses the correlation between the execution time and 
number of transactions to find the
key, so-called distinguisher.
The attacker must be able to do this using a reasonable 
number of samples. Our series of countermeasures increase the number of required samples while keeping and in some cases even improving performance. Here, we first tried to find the number of samples for a successful attack using hardware and software countermeasures separately. Then we used all countermeasures together and tried to attack. As the number of samples increase significantly in this case, a detectable correlation was not observed in a reasonable amount of time. So we calculated the number of samples for this case using statistical analyses.

 The probability of launching an attack can be calculated as
Equation~\ref{eqsa}~\cite{mangard2004hardware},
where $S$ is the number of samples and $\rho_{peak}$ is the highest detectable correlation value.
We have modified this equation by replacing 
$\rho=0$ with $\rho=\rho_{ave}$, because when using a  
smaller number of samples, this value is not negligible. 

\begin{equation} 
\label{eqsa}
\alpha=\Phi\left(\frac{\frac{1}{2}ln\frac{1+\rho_{peak}}{1-\rho_{peak}}-\frac{1}{2}ln\frac{1+\rho_{ave}}{1-\rho_{ave}}}{\sqrt{\frac{2}{S-3}}}\right)
\end{equation} 

Our countermeasures reduce $\rho_{peak}$, then the attacker would need to collect more
timing samples in order to detect $\rho_{peak}$ in the correlation values.
As we discussed earlier, our possible
hardware countermeasure options include randomizing the 
width of coalescing unit, utilizing our hierarchical MHSR design,
both of which can reduce the SNR.
Equation~\ref{eqco} shows the contribution of
noise when computing the correlation value between the
execution and the number of memory transactions.  

\begin{equation} 
\label{eqco0}
{\scriptstyle \rho_{peak}(t_e,n)=\frac{\mu_{t_en}-\mu_{t_e}\mu_{n}}{\sigma_{t_e}\sigma_n}
=\frac{\mu_{(\beta_1n+\beta_0)n+\epsilon
n}-\mu_{\beta_1n+\beta_0}\mu_n-\mu_\epsilon\mu_{n}}{\sigma_{(\beta_1n+\beta_0)}\sigma_n+\sigma_\epsilon\sigma_n}}
\end{equation} 
\begin{equation} 
\label{eqco}
 \rho_{peak}(t_e,n)=\frac{\rho_{peak}(\beta_1n+\beta_0)}{\sqrt{1+\frac{1}{SNR}}}
\end{equation} 

The highest possible value for $\rho_{peak}$ is 1. 
In this case, the variables
would be strongly correlated 
and the measurements would need to be free of any noise. 
When launching a timing attack
on an unprotected GPU, 
the highest
$\rho_{peak}$ that we can obtain is 0.4031, which is limited
only by the inherent noise in the timing channel. The highest value for
$\rho_{peak}$,  
when employing a single (but randomly selected) coalescing 
unit width for each kernel run with average value of 32,
is 0.2994, as shown in Figure~\ref{spacesk}. If we change the coalescing unit width 
dynamically during runtime, the max $\rho_{peak}$ value is 
reduced to 0.0547, meaning that using different sizes for different cache lines 
introduces even more noise. Finally employing our
hierarchical MSHR design, the SNR is reduced to 0.0447. But in 
this case, in order to fully utilize all SMs, we increased the number of 
threads to $32 threads per warp \times 2 warps per SM \times 15 SMs$.

\begin{figure}[h] 
\centering 
\includegraphics[width=0.5\textwidth]{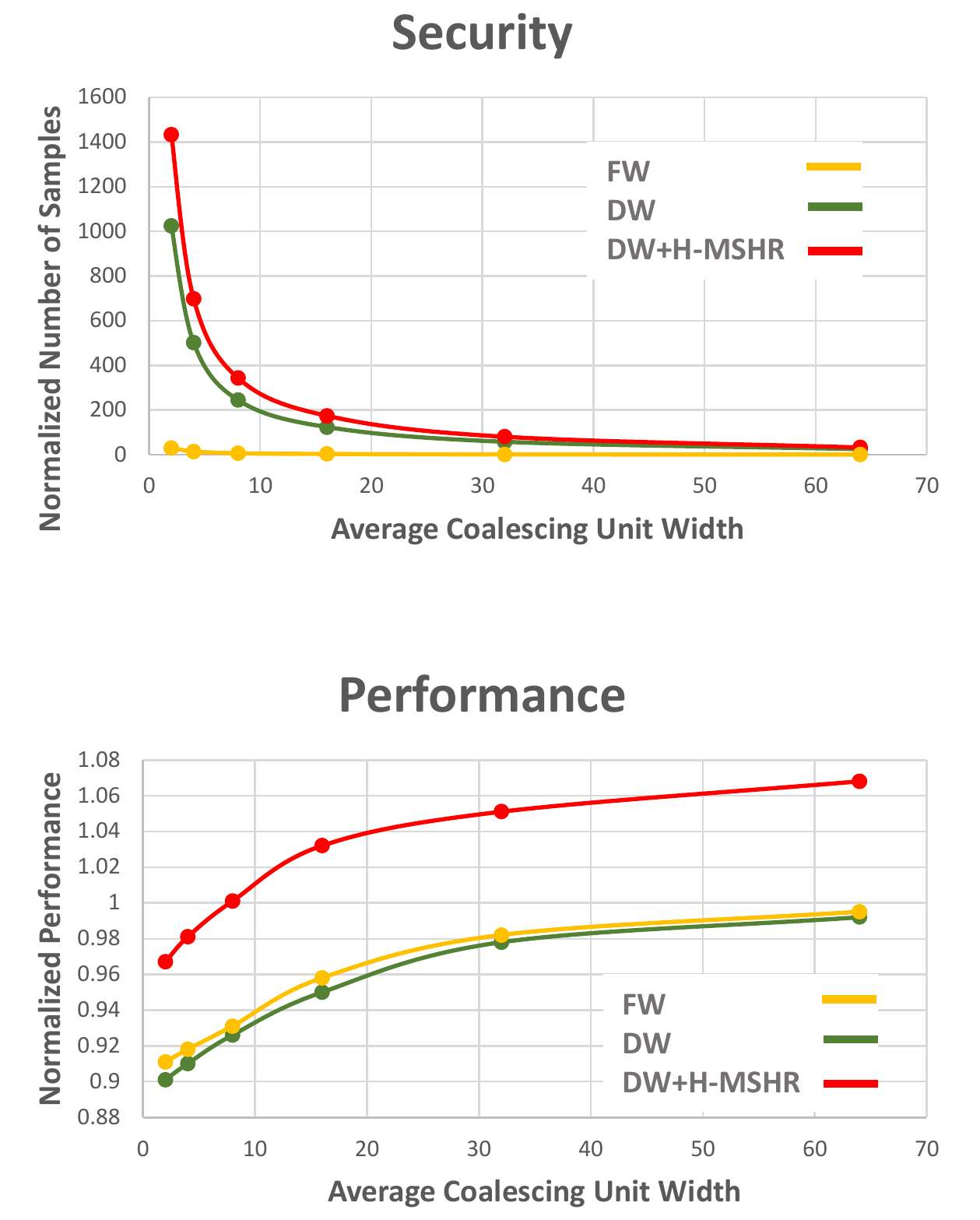}
\caption{Security and performance for hardware countermeasures.FX:Fixed Width, DW:Dynamic Width, H-MSHRs: Hierarchical MSHRs} 
\label{res2h}
\end{figure}

For this series of countermeasures, it is important to choose the right
average width based on the normal distribution of memory access requests in order 
to achieve a good balance between performance and security. Figure~\ref{res2h} 
shows the number of timing samples and performance of different average widths.
Performance is defined as $\frac{1}{execution Time}$. As we can see, using a 
dynamic coalescing width with hierarchical MSHRs improves the resiliency
of the design to attacks, while also boosting performance. With this design,
the effort required by the attacker (measured in the number of timing samples
collected) to launch a successful attack increases by 1433X.  On the other side,
we can enjoy a $7\%$ performance improvement. Considering the fact that deceasing 
the average width makes the design harder to attack, while losing performance, 
the designer can choose different widths based on the performance/security
tradeoff. 
For a design to introduce negligible performance overhead, we should select an average width
no narrower than 16B.  Adopting a value of 16B will increase the difficulty of
launching a successful attack by 175X, while a more aggressive design that increases the
attacker effort by 1433X 
would suffer a $3\%$ loss in performance. 

While hardware countermeasures help to obscure the relationship between the
kernel execution time and the number of transactions by lowering SNR, 
employing our software scheme of rotating T-table columns
will also decrease the correlation since this also changes the
number of transactions. Equation~\ref{eqco3} computes the correlation 
after applying T-table rotation in the original hardware without 
applying hardware countermeasures. In this equation, 
$t$ is the measured time, $n$ is the number of transactions to 
original table, $o$ in the number of transactions after 
rotation and $p_{o=n}$ is the probability of $o=n$. 
In this equation, we assume that the time is not related to $o$
whenever the number of transactions is not equal to $n$.
\begin{equation} 
\label{eqco31}
 \rho_{peak}(t,o)=\frac{p_{o=n}(\mu_{tn}-\mu_{t}\mu_{n})+(1-p_{o=n})(\mu_{t}\mu_{o}-\mu_{t}\mu_{o})}{\sigma_t\sigma_o}
\end{equation} 
\begin{equation} 
\label{eqco3}
\rho_{peak}(t,o)=\rho_{peak}(t,n)p_{o=n}\frac{\sigma_n}{\sigma_o}
\end{equation}  

\begin{figure}[h] 
\centering 
\includegraphics[width=0.5\textwidth]{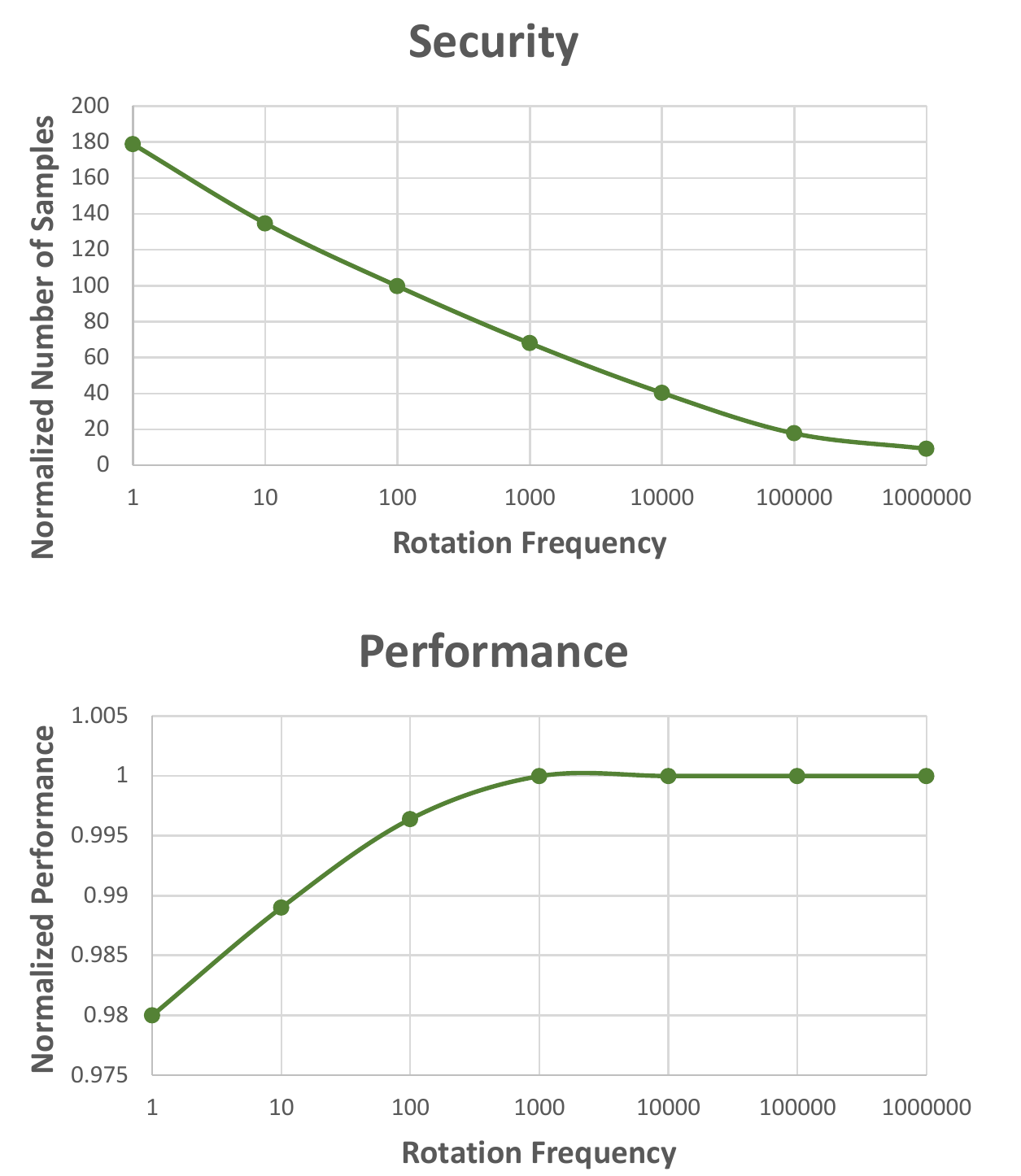}
\caption{Security and performance for software countermeasures.} 
\label{res2s}
\end{figure}

The values for $p_{o=n}$ and $\sigma_o$ depend on the 
frequency of rotations.  Using every 1000 encryptions as the frequency for rotation, 
$\rho_{peak}(t,o)$ can be decreased to $0.0858 \times \rho_{peak}(t,n)$,
without applying any hardware countermeasures. Figure~\ref{res2s} 
shows the resulting security and performance values based on the rotation frequency. 
This figure shows that rotating T-table columns 
every 1M to 1000 samples,
we do not see any detectable performance loss, but
increases the number of samples by 68X. 
Decreasing the time between rotations
to less than 1000 encryptions would result in a
slight performance loss. Using the maximum frequency, 
the T-table will be rotated after every run of the encryption
algorithm, resulting in a 
$2\%$ performance loss, while increasing the effort for
a successful timing attack by 178X.

\begin{figure}[h] 
\centering 
\includegraphics[width=0.5\textwidth]{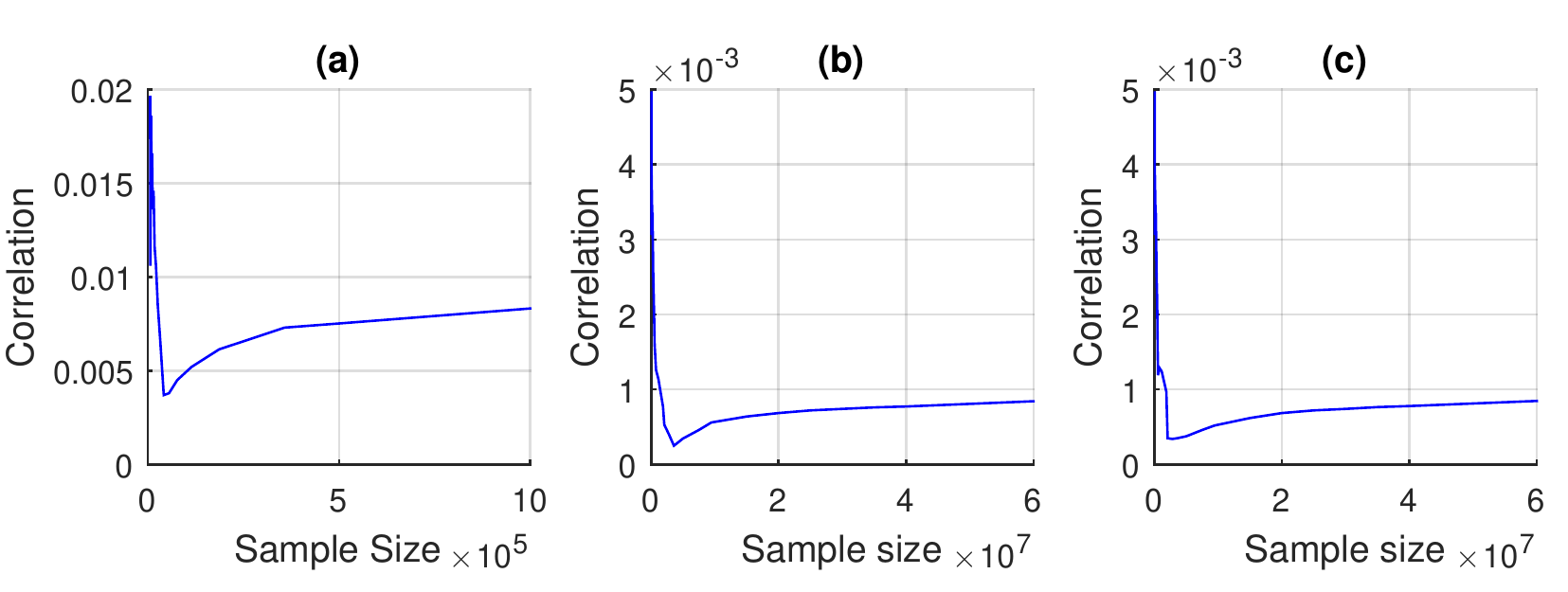}
\caption{$\rho_{peak}$ of $k_5$ vs.
the number of samples a)the baseline attack,
b)the attack only with hardware countermeasures, 
c)the attack only with software countermeasures} 
\label{res1}
\end{figure}

Figure~\ref{res1} shows the correlation values 
when attacking key byte 5,
As we can see, the
number of samples required for the original attack with 
$\alpha=0.9$ is around $5\times10^5$.  When we
apply the hardware (average width=32B )and software (rotation frequency=1000) countermeasures individually,
the number of traces needed grows to
$4\times10^7$ and $3.5\times10^7$, respectively, 
increasing the effort by two orders
of magnitude.

Our hardware and 
software methods reduce the correlation to $0.1109\times p_{peak}$ and $0.0858 \times p_{peak}$,
respectively. When we apply both methods, 
the correlation will be at least 
$0.1109\times 0.0858\times p_{peak}$. By reducing the width of the coalescing unit, 
the number of cache lines accessed will increase,
which will decrease $p_{o=n}$. 
For example, with a 64B L1 cache line, 32 requests will 
access 2 to 16 cache lines. To change each number from this range, 
we have 14 other possible values. But having 32B cache lines 
results in accessing 4 to 64 cache lines. So the possibility of 
hiding the original number of cache lines is higher due 
to using a wider range. Therefore, 
hardware obfuscation will increase the level of obfuscation performed
in software. If the number of required samples with hardware and 
software countermeasures individually are $g_hX$ and $g_sX$, respectively (as compared to the
traces required for the basic attack), the number of 
required traces needed to launch a successful attack using a combination of both
countermeasures will be increased by at least $g_h*g_s$ times.

\vspace{-0.5em}

\section{Conclusion} 
GPUs are vulnerable to timing side-channel attacks.
The memory system of these devices can leak
timing information. If the attacker can time the encyrption, and with 
general knowledge of an encryption algorithm's memory access pattern,
she can recover all bytes of the key. 

In this paper we explore
multiple hardware-based and software-based strategies to 
obfuscate the timing leakage of the memory system on a
GPU.  By randomizing the width of the coalescing unit dynamically,
we are able to introduce significant noise into the timing channel.  
By introducing a second-level
unified MSHRs across all SMs, we can further obfuscate the timing 
channel, while improving encyption performance. We also
propose software-based methods the obfuscate T-table accesses and
the mapping of entries to cache lines. Applying all of these countermeasures
together to thwart an attack, an attacker will still be able to 
recover the encryption key, but
the number of traces required for a successful attack has increased by 
1400*178X over the baseline, while slightly improving AES performance.


\bibliographystyle{ieeetr} \bibliography{ref}

\end{document}